\documentclass[runningheads]{llncs}
\usepackage{epsfig}
\usepackage{graphicx}
\usepackage{amsmath}
\usepackage{amssymb}
\usepackage{booktabs}
\usepackage{multirow}
\usepackage{makecell}
\usepackage{xcolor}
\usepackage{caption}
\makeatletter 
\begin{document}
\title{U-DuDoNet: Unpaired dual-domain network for CT metal artifact reduction}
\titlerunning{U-DuDoNet: Unpaired dual-domain network for CT metal artifact reduction}
	\author{Yuanyuan Lyu\inst{1} \and
	Jiajun Fu\inst{2} \and
	Cheng Peng\inst{3} \and 
	S. Kevin Zhou\inst{4,5,6}}
	\institute{Z$^2$Sky Technologies Inc., Suzhou, China \and 
	Beijing University of Posts and Telecommunications, Beijing, China \and 
	Johns Hopkins University, MD, USA \and 
	Medical Imaging, Robotics, and Analytic Computing Laboratory and Engineering (MIRACLE) Group \and 
    School of Biomedical Engineering \& Suzhou Institute for Advance Research, University of Science and Technology of China, Suzhou 215123, China \and 
    Key Lab of Intelligent Information Processing of Chinese Academy of Sciences (CAS),
    Institute of Computing Technology, CAS, Beijing, 100190, China \email{s.kevin.zhou@gmail.com}}
\maketitle              

\begin{abstract}
Recently, both supervised and unsupervised deep learning methods have been widely applied on the CT metal artifact reduction (MAR) task. Supervised methods such as Dual Domain Network (Du-DoNet) work well on simulation data; however, their performance on clinical data is limited due to domain gap. Unsupervised methods are more generalized, but do not eliminate artifacts completely through the sole processing on the image domain. To combine the advantages of both MAR methods, we propose an unpaired dual-domain network (U-DuDoNet) trained using unpaired data. Unlike the artifact disentanglement network (ADN) that utilizes multiple encoders and decoders for disentangling content from artifact, our U-DuDoNet directly models the artifact generation process through additions in both sinogram and image domains, which is theoretically justified by an additive property associated with metal artifact. Our design includes a self-learned sinogram prior net, which provides guidance for restoring the information in the sinogram domain, and cyclic constraints for artifact reduction and addition on unpaired data. Extensive experiments on simulation data and clinical images demonstrate that our novel framework outperforms the state-of-the-art unpaired approaches.
\end{abstract}

\section{Introduction}
Computed tomography (CT) reveals the underlying anatomical structure within the human body. However, when a metallic object is present, metal artifacts appear in the image because of beam hardening, scatters, photon starvation, etc.~\cite{barrett2004artifacts,boas2012ct,meyer2010normalized}, degrading the image quality and limiting its diagnostic value.

With the success of deep learning in medical image processing~\cite{zhou2021review,zhou2019handbook}, deep learning has been used for metal artifact reduction (MAR). Single-domain networks~\cite{ghani2019fast,wang2018conditional,cnnmar} have been proposed to address MAR with success. Lin \textit{et al.} are the first to introduce dual-domain network (DuDoNet) to reduce metal artifacts in the sinogram and image domain jointly and DuDoNet shows further advantages over single-domain networks and traditional approaches~\cite{chang2018prior,jin2015model,kalender1987reduction,karimi2015metal,meyer2010normalized}. Following this work, variants of the dual-domain architecture~\cite{lyu2020encoding,wang2021dan,yu2020deep} have been designed. However, all the above-mentioned networks are supervised and rely on paired clean and metal-affected images. Since such clinical data is hard to acquire, simulation data are widely used in practice. Thus, supervised models may over-fit to simulation and \textit{do not generalize well to real clinical data}.

Learning from unpaired, real data is thus of interest. To this, Liao \textit{et al.} propose ADN~\cite{adn2019_tmi,adn2019_miccai}, which separates content and artifact in the latent spaces with multiple encoders and decoders and induces unsupervised learning via various forms of image generation and specialized loss functions. An artifact consistency loss is introduced to retain anatomical preciseness during MAR. The loss is based on the assumption that metal artifacts are additive.
Later on, Zhao \textit{et al.}~\cite{zhao2020unsupervised} design a simple reused convolutional network (RCN) of encoders and decoders to recurrently generating both artifact and non-artifact images. RCN also adopts the additive metal artifacts assumption. However, \textit{neither of the works has theoretically proved the property}. Moreover, without the aid of processing in sinogram domain, both methods have \textit{limited effect on removing strong metal artifacts}, such as the dark and bright bands around metal objects. 

In this work, we analytically derive the additive property associated with metal artifacts and propose an unpaired dual-domain MAR network (U-DuDoNet). Without using complicated encoders and decoders, our network \textit{directly estimates} the additive component of metal artifacts, jointly using two U-Nets on two domains: a sinogram-based estimation net (S-Net) and an image-based estimation net (I-Net). S-Net first restores sinogram data and I-net removes additional streaky artifacts. Unpaired learning is achieved with cyclic artifact reduction and synthesis processes. Strong metal artifacts can be reduced in the sinogram domain with prior knowledge. Specifically, sinogram enhancement is guided by a self-learned sinogram completion network (P-Net) with clean images. Both simulation and clinical data show our method outperforms competing unsupervised approaches and has better generalizability than supervised approaches.

\section{Additive Property for Metal Artifacts} \label{sec:property}
Here, we prove metal artifacts are inherently additive up to mild assumptions. The CT image intensity represents the attenuation coefficient. Let $ X^c(E) $ be a normal attenuation coefficient image at energy level $E$. In a polychromatic x-ray system, the ideal projection data (sinogram) $ S^c $ can be expressed as,
\begin{equation}
S^c = -ln\int \eta(E)e^{-\mathcal{P}(X^c(E))}d E, 
\end{equation}
where $ \mathcal{P} $ and $\eta(E)$ denote forward projection (FP) operator and  fractional energy at $E$. Comparing with metal, the attenuation coefficient of normal body tissue is almost constant with respect to $E$, thus we have $X^c = X^c(E)$ and $S^c = \mathcal{P}(X^c)$. Without metal, filtered back projection (FBP) operator $\mathcal{P}^*$ provides a clean CT image $I^c$ as a good estimation of $X^c$, 
$I^c=\mathcal{P}^*(S^c) = \mathcal{P}^*(\mathcal{P}(X^c))$.

Metal artifacts appear mainly because of beam hardening. An attenuation coefficient image with metal $ X^a(E)$ can be split into a relatively constant image without metal $X^{ac}$ and a metal-only image $X^m(E)$ varies rapidly against $E$, $X^a(E)=X^{ac}+ X^m(E)$. Often $X^m(E)$ is locally constrained. The contaminated sinogram $S^a$ can be given as,
\begin{equation}
S^a = -ln\int \eta(E)e^{-\mathcal{P}(X^a(E))}d E = \mathcal{P}(X^{ac})-ln\int \eta(E)e^{-\mathcal{P}(X^m(E))}d E.
\end{equation}
And the reconstructed metal-affected CT image $I^a$ is,
\begin{equation} \label{eq_ma_ct}
I^a = \mathcal{P}^*(\mathcal{P}(X^{ac}))-\mathcal{P}^*(ln\int \eta(E)e^{-\mathcal{P}(X^m(E))}d E)= I^{ac}+F(X^m(E)). 
\end{equation}
Here, $\mathcal{P}^*(\mathcal{P}(X^{ac}))$ is the MAR image $I^{ac}$ and the second term introduces streaky and band artifacts, which is a function of $X^m(E)$. Since metal artifacts are caused only by $X^m(E)$, we can create a plausible artifact-affected CT $I^{ca}$ by adding the artifact term to an arbitrary, clean CT image: $I^{ca} = I^c + F(X^m(E))$.

\section{Methodology}
\begin{figure*}[t]
	\begin{center}
		\includegraphics[width=0.9\linewidth]{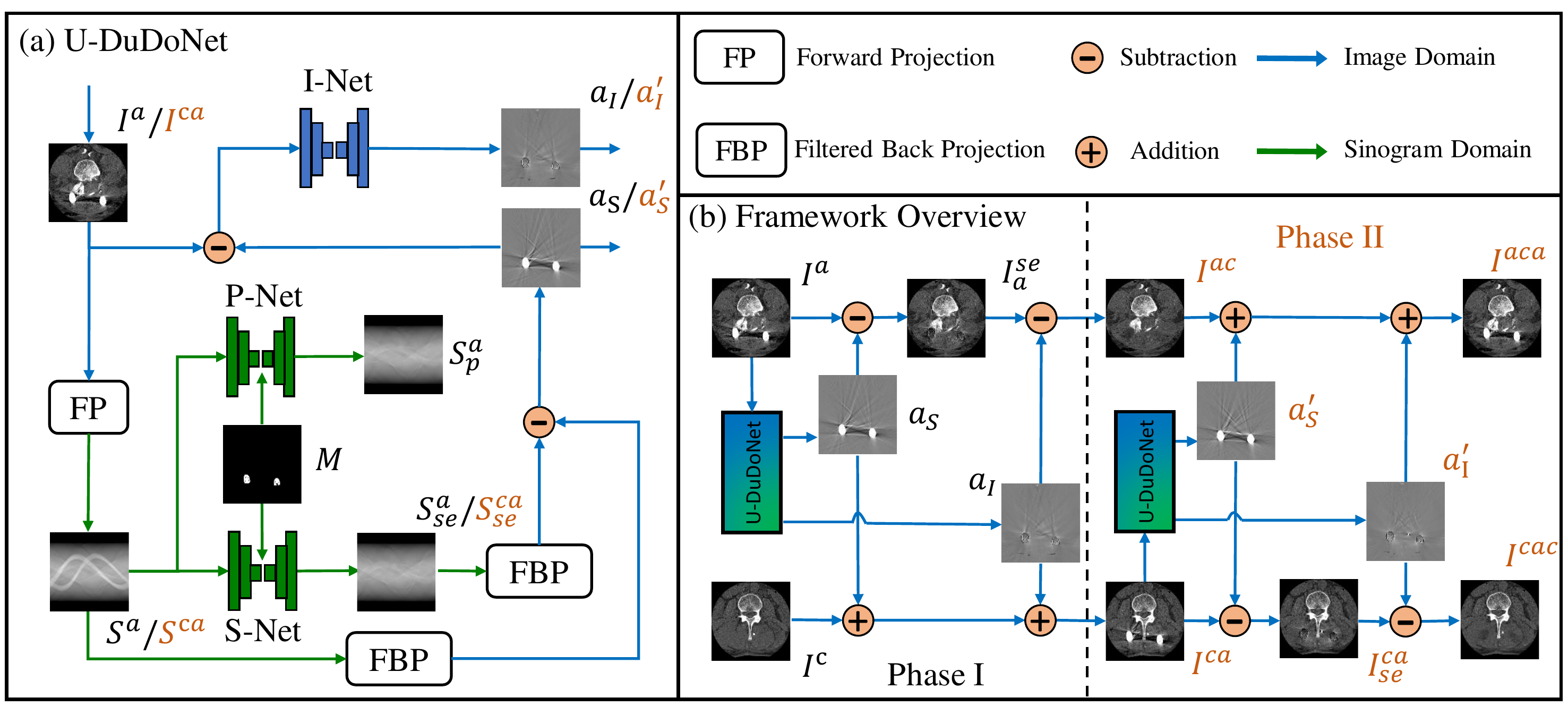}
	\end{center}
	\caption{ (a) Unpaired dual-domain network (U-DuDoNet). The input is either a real artifact-affected image $I^a$, or a synthetic artifact-affected image $I^{ca}$. (b) The proposed cyclical MAR framework. The notations in Phase \uppercase\expandafter{\romannumeral1} and Phase \uppercase\expandafter{\romannumeral2} are marked with black and brown, respectively.}
	\label{fig:network}
\end{figure*}

Fig.~\ref{fig:network}b shows the proposed cyclical MAR framework. In Phase \uppercase\expandafter{\romannumeral1}, our framework first estimates artifact components $a_{S}$ and $a_{I}$ through U-DuDoNet from $I^a$, see Fig.~\ref{fig:network}a. In Phase \uppercase\expandafter{\romannumeral2}, based on the additive metal artifact property (Section~\ref{sec:property}), plausible clean image $I^{ac}$ and metal-affected image $I^{ca}$ could be generated, $ I^{ac} = I^a - a_{S} - a_{I}$, $I^{ca} = I^c + a_{S} + a_{I}$. Then, the artifact components should be removable from $I^{ca}$ by U-DuDoNet, resulting in $a_{S}^{\prime}$ and $a_{I}^{\prime}$. In the end, reconstructed images $I^{aca}$, $I^{cac}$ can be obtained through subtracting or adding the artifact components, $I^{aca} = I^{ac} + a_{S}^{\prime} + a_{I}^{\prime}$ , $I^{cac} = I^{ca} - a_{S}^{\prime} - a_{I}^{\prime}$. 

\subsection{Network Architecture}
\textbf{Artifact component estimation in sinogram domain.}
Strong metal artifacts like dark and bright bands can not be suppressed completely by image domain processing, while metal artifacts are inherently local in the sinogram domain. Thus, we aim to reduce metal shadows by sinogram enhancement.  

First, we acquire metal corrupted sinograms ($S^a$ and $S^{ca}$) by forward projecting $I^a$ and $I^{ca}$: $S^a = \mathcal{P}(I^a), S^{ca}=\mathcal{P}(I^{ca})$. Then, we use a pre-trained prior net (P-Net) to guide the sinogram restoration process. P-Net is an inpainting net $\phi_{P}$ that treats the metal-affected area in sinogram as missing and aims to complete it, i.e., $S^a_p = \phi_{P}(S^a, M_t)$. Here $M_t$ denotes a binary metal trace, $M_t = \delta(\mathcal{P}(M))$, where $M$ is a metal mask and $\delta(\cdot)$ is a binary indicator function. We adopt a mask pyramid U-Net as $\phi_{P}$ from~\cite{liao2019generative,lin2019dudonet}. To train $\phi_{P}$, we artificially inject masks into clean sinograms. 

Then, we use a sinogram network (S-Net) to predict enhanced sinogram $S^a_{se}$, $S^{ca}_{se}$ from $S^a$, $S^{ca}$, respectively.
\begin{equation}
S^a_{se} = \phi_{S}(S^a, \mathcal{P}(M))\odot(1-M_t)+S^a, S^{ca}_{se} = \phi_{S}(S^{ca}, \mathcal{P}(M))\odot(1-M_t)+S^{ca},
\end{equation}
where $\phi_{S}$ represents a U-Net\cite{ronneberger2015u} of depth 2. Residual learning~\cite{he2016deep} is applied to ease the training process, and singoram prediction is limited to the $M_t$ region. To prevent information loss from discrete operators, we obtain the sinogram artifact component as a difference image between reconstructed input image and reconstructed enhanced sinogram,
\begin{equation}
a_{S} = \mathcal{P}^{*}(S^a) -\mathcal{P}^{*}(S^a_{se}), a_{S}^{\prime} = \mathcal{P}^{*}(S^{ca}) -\mathcal{P}^{*}(S^{ca}_{se}).
\end{equation}

\noindent\textbf{Artifact component estimation in image domain.}
As sinogram data inconsistency leads to secondary artifacts in the whole image, we further use an image domain network (I-Net) to reduce newly introduced and other streaky artifacts. Let $\phi_{I}$ denote I-Net, which is a 5-depth U-Net. First, sinogram enhanced images are obtained by subtracting sinogram artifact component from corrupted images, $I^a_{se} = I^a - a_{S}, I^{ca}_{se} = I^{ca}-a_{S}^{\prime}$. Then, I-Net takes a sinogram enhanced image, and outputs an artifact component in image domain ($a_{I}$ or $a_{I}^{\prime}$),
\begin{equation}
a_{I} = \phi_{I}(I^a_{se}), a_{I}^{\prime} = \phi_{I}(I^{ca}_{se}).
\end{equation}

\subsection{Dual-domain Cyclic Learning}
To obviate the need of paired data, we use cycle loss and artifact consistency loss as cyclic MAR constraints and adopt adversarial loss. Besides, we take advantage of prior knowledge to guide the data restoration in Phase \uppercase\expandafter{\romannumeral1} and apply dual-domain loss to encourage the data fidelity in Phase \uppercase\expandafter{\romannumeral2}. 

\noindent\textbf{Cycle loss.} By cyclic artifact reduction and synthesis, the original and reconstructed images should be identical. We use $L_1$ loss to minimize the distance,
\begin{equation}
\mathcal{L} _{cycle} = || I^a - I^{aca}||_1 + || I^c - I^{cac}||_1.
\end{equation}
\textbf{Artifact consistency loss.} To ensure the artifacts components added to $I^c$ could be removed completely when applying the same network on $I^{ca}$, the artifact components estimated from $I^a$ and $I^{ca}$ should be the same,
\begin{equation}
\mathcal{L} _{art} = || a_{S} - a_{S}^{\prime}||_1 + || a_{I} - a_{I}^{\prime}||_1.
\end{equation}
\textbf{Adversarial loss.} The synthetic images, $I^{ca}$ and $I^{ac}$, should be indistinguishable to input images. Since paired groundtruth is not available, we adopt PatchGAN~\cite{isola2017image} as discriminators $D^a$ and $D^c$ to apply adversarial learning. Since metal affected images always contain streaks, we add gradient image generated by Sobel operator $\bigtriangledown$ as an additional channel of the input of $D^a$ and $D^c$ to achieve better performance. The loss would be written as,
\begin{equation}
\begin{aligned}
\mathcal{L}_{adv} &= \mathbb{E}[\log D^a(I^a, \bigtriangledown I^a)] + \mathbb{E}[1- \log D^a(I^{ca}, \bigtriangledown I^{ca})] \\ & + \mathbb{E}[\log D^c(I^c, \bigtriangledown I^c)] + \mathbb{E}[1- \log D^c(I^{ac}, \bigtriangledown I^{ac})].
\end{aligned}
\end{equation}
\textbf{Fedility loss.} To learning artifact reduction from generated $I^{ca}$, we minimize the distances between $S^{ca}_{se}$ and $S^c$, $I^{ca}_{se}$ and $I^c$, 
\begin{equation}
\mathcal{L} _{fed} = || S^{ca}_{se}- S^c||_1 + || I^{ca}_{se} - I^c||_1.
\end{equation}
\textbf{Prior loss.} 
Inspried by DuDoNet~\cite{lin2019dudonet}, sinogram inpainting network provides smoothed estimation of sinogram data within $M_t$. Thus, we use a Gaussian blur operation $\mathcal{G}_{\sigma_s}$ with a scale of $\sigma_s$ and $L_2$ loss to minimize the distance between blurred prior and enhanced sinogram.  Meanwhile, inspired by~\cite{jin2019unsupervised}, blurred sinogram enhanced image also serves as an good estimation of blurred MAR image. Also, we minimize the distance between low-pass versions of sinogram enhanced and MAR images with a Gaussian blur operation $\mathcal{G}_{\sigma_i}$ to stabilize the unsupervised training. The prior loss could be formulated as,
\begin{equation}
\mathcal{L} _{prior} = || \mathcal{G}_{\sigma_s}(S^a_p) - \mathcal{G}_{\sigma_s}(S^a_{se})||_2 + || \mathcal{G}_{\sigma_i}(I^a_{se}) - \mathcal{G}_{\sigma_i}(I^{ac})||_2.
\end{equation}

The overall objective function is the weighted sum of all the above losses, we empirically set the weight of $\mathcal{L}_{adv}$ to 1, and the weights of $\mathcal{L} _{prior}$, $\mathcal{L} _{art}$ to 10, and the weights of the other losses to 100. We set $\sigma_s$ to 1 and $\sigma_i$ to 3 in $\mathcal{L} _{prior}$.

\section{Experiment}
\subsection{Experimental Setup}
\textbf{Datasets.} Following~\cite{adn2019_tmi}, we evaluate our model on both simulation and clinical data. For simulation data, we generate images with metal artifacts using the method in~\cite{cnnmar}. From DeepLesion~\cite{yan2018deep}, we randomly choose 3,984 clean images combining with 90 metal masks for training and additional 200 clean images combining with 10 metal masks for testing. For unsupervised training, we spilt 3,984 images into two groups and randomly select one metal corrupted image and one clean image. For clinical data, we select 6,146 images with artifacts and 21,002 clean images for training from SpineWeb~\cite{glocker2012automatic,glocker2013vertebrae}. Additional 124 images with metal are used for testing.

\noindent\textbf{Implementation and metrics.}
We implement our model with the PyTorch framework and differential FP and FBP operators with ODL library~\cite{adler2017operator}. We train the model for 50 epochs using an Adam optimizer with a learning rate of $1\times10^{-4}$ and a batch size of 2. For clinical data, we train another model using unpaired images for 20 epochs and the metal mask is segmented with a threshold of 2,500 HU. We use peak signal-to-noise ratio (PSNR) and structural similarity index (SSIM) to evaluate the corrected image. 
 
\noindent\textbf{Baselines.}
We compare U-DuDoNet with multiple state-of-the-art (SOTA) MAR methods. DuDoNet~\cite{lin2019dudonet}, DuDoNet++~\cite{lyu2020encoding}, DSCIP~\cite{yu2020deep} and DAN-Net~\cite{wang2021dan} are supervised methods which are trained with simulation data and tested on both simulation and clinical data. DuDoNet, DuDoNet++ and DAN-Net share the same SE-IE architecture with an image enhancement (IE) following sinogram enhancement (SE) network, while DSCIP adopts an IE-SE architecture that predicts a prior image first then outputs the sinogram enhanced image. RCN~\cite{zhao2020unsupervised} and ADN~\cite{adn2019_tmi} are unsupervised and can be trained and tested on each dataset.

\begin{table}[t]
	\begin{center}
		\resizebox{0.9\linewidth}{!}{
			\begin{tabular}{l|l|c|c|c}
				\toprule[1pt]
				&\multirow{2}{*}{Method} & \footnotesize{Sinogram domain CT}  &  \footnotesize{Image domain CT} &\multirow{2}{*}{Running time (ms)} \\
				&&PSNR(dB)/SSIM&PSNR(dB)/SSIM& \\
				\hline
				\hline
				Metal & n/a& n/a & 27.23/0.692 & n/a \\
				\hline
				\multirow{4}{*}{Supervised}&DuDoNet~\cite{lin2019dudonet} & 32.20/0.755 & 36.95/0.927 & 60.38\\  
				&DuDoNet++~\cite{lyu2020encoding} & 32.20/0.751 & 37.65/\textbf{0.953} & 39.77\\ 
				&DSCIP~\cite{yu2020deep} & 29.22/0.624 & 30.06/0.790 &  62.01\\  
				&DAN-Net~\cite{wang2021dan}& 32.48/0.752 & \textbf{39.73}/0.944 & 63.75\\ 
				\hline
				\multirow{3}{*}{Unsupervised}&RCN~\cite{zhao2020unsupervised} & n/a &32.98/0.918& 38.18 \\
				&ADN~\cite{adn2019_tmi} & n/a & 33.81/0.926 & 37.66\\
				&U-DuDoNet (ours) & 30.47/0.722 & \textbf{34.54}/\textbf{0.934} & 63.59 \\
				\toprule[1pt]
			\end{tabular}
		}
	\end{center}
	\caption{Quantitative comparison of different SOTA methods.}
	\label{table:sota} 
	 \vspace{-0.2in}
\end{table}

\begin{figure*}[t]
	\begin{center}
		\scriptsize
		\begin{minipage}{0.9\textwidth}
			\centering
			\begin{minipage}[t]{0.19\textwidth}
				\centering
				\includegraphics[width=1\textwidth]{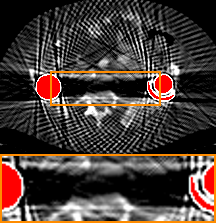}
				Corrupted CT
			\end{minipage}
			\begin{minipage}[t]{0.19\textwidth}
				\centering
				\includegraphics[width=1\textwidth]{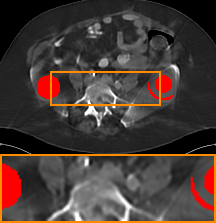}
				DuDoNet
			\end{minipage}
			\begin{minipage}[t]{0.19\textwidth}
				\centering
				\includegraphics[width=1\textwidth]{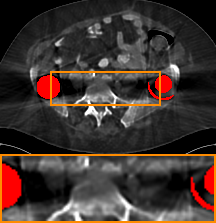}
				DuDoNet++
			\end{minipage}
			\begin{minipage}[t]{0.19\textwidth}
				\centering
				\includegraphics[width=1\textwidth]{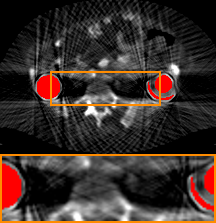}
				DSCIP
			\end{minipage}
			\begin{minipage}[t]{0.19\textwidth}
				\centering
				\includegraphics[width=1\textwidth]{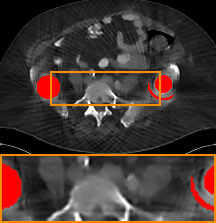}
				DAN-Net
			\end{minipage}
		\end{minipage}
		
		\begin{minipage}{0.9\textwidth}
			\centering
			\begin{minipage}[t]{0.19\textwidth}
				\centering
				\includegraphics[width=1\textwidth]{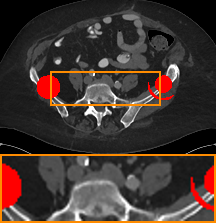}
				Groundtruth
			\end{minipage}
			\begin{minipage}[t]{0.19\textwidth}
				\centering
				\includegraphics[width=1\textwidth]{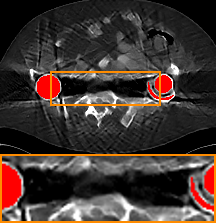}
				RCN
			\end{minipage}
			\begin{minipage}[t]{0.19\textwidth}
				\centering
				\includegraphics[width=1\textwidth]{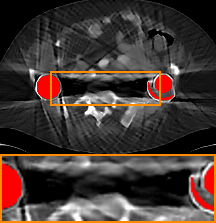}
				ADN
			\end{minipage}
			\begin{minipage}[t]{0.19\textwidth}
				\centering
				\includegraphics[width=1\textwidth]{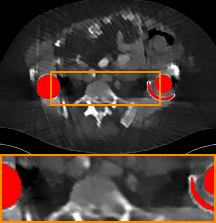}
				Ours
			\end{minipage}
		\end{minipage}
	\end{center}
	\caption{Visual comparisons with the SOTA methods on simulation data. The display window is [-200, 600] HU and red pixels stand for metal implants.}
	\label{fig:sota_simulation}
\end{figure*}

\subsection{Comparison on Simulated and Real Data}

\noindent\textbf{Simulated Data.}
From Table~\ref{table:sota}, we observe that DAN-Net achieves the highest PSNR and DuDoNet++ achieves the highest SSIM. All methods with SE-IE architecture outperform DSCIP. The reason is image enhancement network helps recover details and bridge the gap between real images and reconstructed images. Among all the unsupervised methods, our model attains the best performance, with an improvement of 0.73 dB in PSNR compared with ADN. Besides, our model runs as fast as the supervised dual-domain models but slower than image-domain unsupervised models.
Figure~\ref{fig:sota_simulation} shows the visual comparisons of a case. The zoomed subfigure shows that metallic implants induce dark bands in the region between two implants or along the direction of dense metal pixels. Learning from linearly interpolated (LI) sinogram, DuDoNet removes the dark bands and streaky artifacts completely but smooths out the details around the metal. DuDoNet++ and DSCIP could not remove the dark bands completely as they learn from the corrupted images. DAN-Net contains fewer streaks than DuDoNet++ and DSCIP since it recovers from a blended sinogram of LI and metal-affected data. Among all the unsupervised methods, only our model recovers the bony structure in dark bands and contains least streaks.

\begin{figure*}[htpb]
	\begin{center}
		\scriptsize
		\begin{minipage}{0.9\textwidth}
			\centering
			\begin{minipage}[t]{0.24\textwidth}
				\centering
				\includegraphics[width=1\textwidth]{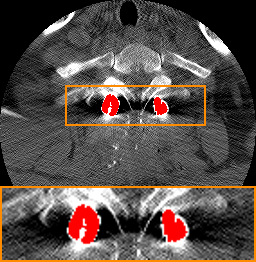}
				Corrupted CT
			\end{minipage}
			\begin{minipage}[t]{0.24\textwidth}
				\centering
				\includegraphics[width=1\textwidth]{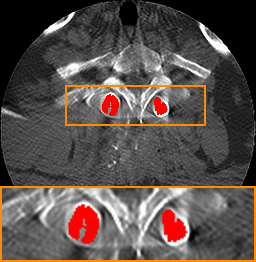}
				DuDoNet
			\end{minipage}
			\begin{minipage}[t]{0.24\textwidth}
				\centering
				\includegraphics[width=1\textwidth]{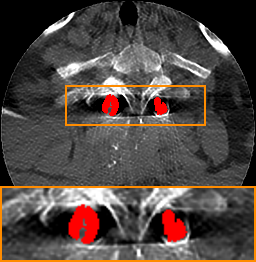}
				DuDoNet++
			\end{minipage}
			\begin{minipage}[t]{0.24\textwidth}
				\centering
				\includegraphics[width=1\textwidth]{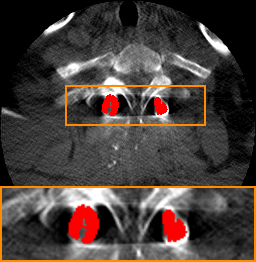}
				DSCIP
			\end{minipage}
		\end{minipage}
				\begin{minipage}{0.9\textwidth}
			\centering
			\begin{minipage}[t]{0.24\textwidth}
				\centering
				\includegraphics[width=1\textwidth]{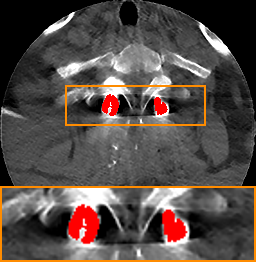}
				DAN-Net
			\end{minipage}
			\begin{minipage}[t]{0.24\textwidth}
				\centering
				\includegraphics[width=1\textwidth]{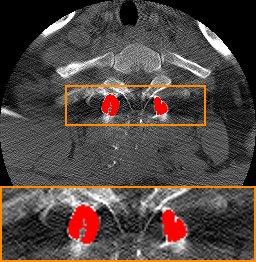}
				RCN
			\end{minipage}
			\begin{minipage}[t]{0.24\textwidth}
				\centering
				\includegraphics[width=1\textwidth]{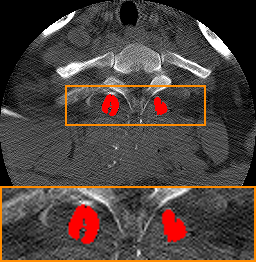}
				ADN
			\end{minipage}
			\begin{minipage}[t]{0.24\textwidth}
				\centering
				\includegraphics[width=1\textwidth]{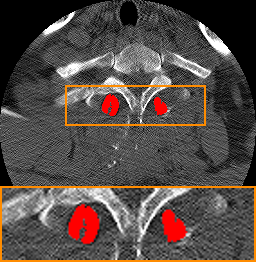}
				Ours
			\end{minipage}
		\end{minipage}
	\end{center}
	
	\caption{Visual comparisons with the SOTA methods on clinical data.}
	\label{fig:sota_spineweb}
\end{figure*}

\noindent\textbf{Real Data.}
Fig.~\ref{fig:sota_spineweb} shows a clinical CT image with two rods on each side of the spinous process of a vertebra. The implants induce severe artifacts, which make some bone part invisible. DuDoNet recovers the bone but introduces strong secondary artifacts. DuDoNet++, DSCIP, and DAN-Net do not generalize to clinical data as the dark band remains in the MAR images. Besides, all the supervised methods output smoothed images as training images from DeepLesion might be reconstructed by a soft tissue kernel. The MAR images of the unsupervised method could retain the sharpness of the original image. But, RCN and ADN do not reduce the artifacts completely or retain the integrity of bone structures near the metal as these structures might be confused with artifacts. Our model removes the dark band while retaining the structures around the rods. More visual comparisons are in the supplemental material.

\begin{table}[t]
	\begin{center}
		\resizebox{0.9\linewidth}{!}{
			\begin{tabular}{l|c|c|c}
				\toprule[1pt]
				PSNR(dB)/SSIM & M1 (Image Domain) & M2 (Dual-domain) & M3 (Dual-domain $+$ prior)  \\
				\hline
				\hline
				$I^{a}_{se}$    &  n/a     &29.99/0.730     &  30.47/0.730     \\
				$I^{ca}$        & 32.97/0.927  & 33.30/0.930    &  \textbf{34.54}/\textbf{0.934} \\
				\toprule[1pt]
			\end{tabular}
		}
	\end{center}
	\caption{Quantitative comparison of different variants of our model.}
	\label{table:ablation} 
	\vspace{-0.2in}
\end{table}

\begin{figure*}[t]
	\centering
	\scriptsize
	\begin{minipage}{0.9\textwidth}
		\begin{minipage}[t]{0.195\textwidth}
			\centering
			\includegraphics[width=1\textwidth]{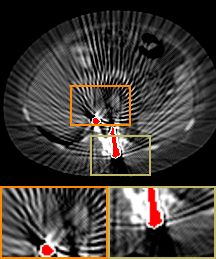}
			Corrupted CT
		\end{minipage}
		\begin{minipage}[t]{0.195\textwidth}
			\centering
			\includegraphics[width=1\textwidth]{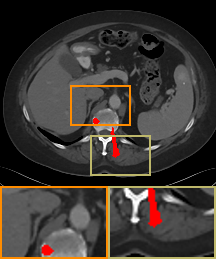}
			Groundtruth
		\end{minipage}
		\begin{minipage}[t]{0.195\textwidth}
			\centering
			\includegraphics[width=1\textwidth]{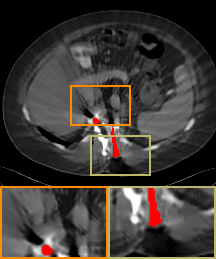}
			M1
		\end{minipage}
		\begin{minipage}[t]{0.195\textwidth}
			\centering
			\includegraphics[width=1\textwidth]{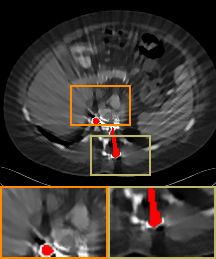}
			M2
		\end{minipage}
		\begin{minipage}[t]{0.195\textwidth}
			\centering
			\includegraphics[width=1\textwidth]{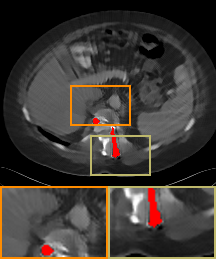}
			M3(full)
		\end{minipage}
	\end{minipage}
	\caption{Visual comparisons of different variants of our model.}
	\label{fig:ablation}
\end{figure*}

\subsection{Ablation Study}
We evaluate the effectiveness of different components in our full model. Table~\ref{table:ablation} shows the configuration of our ablation models. Briefly, M1 refers to the model with I-Net and $\mathcal{L} _{cycle}$, $\mathcal{L} _{adv}$, M2 refers to M1 plus S-Net, $\mathcal{L} _{gt}$, $\mathcal{L} _{art}$, and M3 refers to M2 plus P-Net, $\mathcal{L} _{prior}$. As shown in Table~\ref{table:ablation} and Fig.~\ref{fig:ablation}, M1 has the capability of MAR in image domain, but strong artifacts like dark bands and streaks remain in the output image. Dual-domain learning increases the PSNR by 0.33 dB, and the dark bands are partial removed in the corrected image, but streaks show up as sinogram enhancement might be not perfect. With the aid of prior knowledge, M3 could remove the dark bands completely and further suppresses the secondary artifacts.

\section{Conclusion}
In this paper, we present an unpaired dual-domain network (U-DuDoNet) that exploits the additive property of artifact modeling for metal artifact reduction. In particular, we first remove the strong metal artifacts in sinogram domain and then suppress the streaks in image domain. Unsupervised learning is achieved via cyclic additive artifact modeling, i.e. we try to remove the same artifact after inducing artifact in an unpaired clean image. We also apply prior knowledge to guide data restoration. Qualitative evaluations and visual comparisons demonstrate that our model yields better MAR performance than competing methods. Moreover, our model shows great potential when applied to clinical images.

\newpage
\bibliographystyle{splncs04}
\bibliography{egbib}

\begin{thebibliography}{10}
\providecommand{\url}[1]{\texttt{#1}}
\providecommand{\urlprefix}{URL }
\providecommand{\doi}[1]{https://doi.org/#1}

\bibitem{adler2017operator}
Adler, J., Kohr, H., Oktem, O.: Operator discretization library (odl). Software
  available from https://github. com/odlgroup/odl  (2017)

\bibitem{barrett2004artifacts}
Barrett, J.F., Keat, N.: Artifacts in ct: recognition and avoidance.
  Radiographics  \textbf{24}(6),  1679--1691 (2004)

\bibitem{boas2012ct}
Boas, F.E., Fleischmann, D.: Ct artifacts: causes and reduction techniques.
  Imaging in Medicine  \textbf{4}(2),  229--240 (2012)

\bibitem{chang2018prior}
Chang, Z., Ye, D.H., Srivastava, S., Thibault, J.B., Sauer, K., Bouman, C.:
  Prior-guided metal artifact reduction for iterative x-ray computed
  tomography. IEEE transactions on medical imaging  \textbf{38}(6),  1532--1542
  (2018)

\bibitem{ghani2019fast}
Ghani, M.U., Karl, W.C.: Fast enhanced ct metal artifact reduction using data
  domain deep learning. IEEE Transactions on Computational Imaging  (2019)

\bibitem{glocker2012automatic}
Glocker, B., Feulner, J., Criminisi, A., Haynor, D.R., Konukoglu, E.: Automatic
  localization and identification of vertebrae in arbitrary field-of-view ct
  scans. In: International Conference on Medical Image Computing and
  Computer-Assisted Intervention. pp. 590--598. Springer (2012)

\bibitem{glocker2013vertebrae}
Glocker, B., Zikic, D., Konukoglu, E., Haynor, D.R., Criminisi, A.: Vertebrae
  localization in pathological spine ct via dense classification from sparse
  annotations. In: International Conference on Medical Image Computing and
  Computer-Assisted Intervention. pp. 262--270. Springer (2013)

\bibitem{he2016deep}
He, K., Zhang, X., Ren, S., Sun, J.: Deep residual learning for image
  recognition. In: Proceedings of the IEEE conference on computer vision and
  pattern recognition. pp. 770--778 (2016)

\bibitem{isola2017image}
Isola, P., Zhu, J.Y., Zhou, T., Efros, A.A.: Image-to-image translation with
  conditional adversarial networks. In: Proceedings of the IEEE conference on
  computer vision and pattern recognition. pp. 1125--1134 (2017)

\bibitem{jin2015model}
Jin, P., Bouman, C.A., Sauer, K.D.: A model-based image reconstruction
  algorithm with simultaneous beam hardening correction for x-ray ct. IEEE
  Transactions on Computational Imaging  \textbf{1}(3),  200--216 (2015)

\bibitem{jin2019unsupervised}
Jin, X., Chen, Z., Lin, J., Chen, Z., Zhou, W.: Unsupervised single image
  deraining with self-supervised constraints. In: 2019 IEEE International
  Conference on Image Processing (ICIP). pp. 2761--2765. IEEE (2019)

\bibitem{kalender1987reduction}
Kalender, W.A., Hebel, R., Ebersberger, J.: Reduction of ct artifacts caused by
  metallic implants. Radiology  \textbf{164}(2),  576--577 (1987)

\bibitem{karimi2015metal}
Karimi, S., Martz, H., Cosman, P.: Metal artifact reduction for ct-based
  luggage screening. Journal of X-ray science and technology  \textbf{23}(4),
  435--451 (2015)

\bibitem{adn2019_tmi}
{Liao}, H., {Lin}, W., {Zhou}, S.K., {Luo}, J.: Adn: Artifact disentanglement
  network for unsupervised metal artifact reduction. IEEE Transactions on
  Medical Imaging  (2019). \doi{10.1109/TMI.2019.2933425}

\bibitem{liao2019generative}
Liao, H., Lin, W.A., Huo, Z., Vogelsang, L., Sehnert, W.J., Zhou, S.K., Luo,
  J.: Generative mask pyramid network for ct/cbct metal artifact reduction with
  joint projection-sinogram correction. In: International Conference on Medical
  Image Computing and Computer-Assisted Intervention. pp. 77--85. Springer
  (2019)

\bibitem{adn2019_miccai}
Liao, H., Lin, W.A., Yuan, J., Zhou, S.K.Z., Luo, J.: Artifact disentanglement
  network for unsupervised metal artifact reduction. In: International
  Conference on Medical Image Computing and Computer-Assisted Intervention
  (MICCAI) (2019)

\bibitem{lin2019dudonet}
Lin, W.A., Liao, H., Peng, C., Sun, X., Zhang, J., Luo, J., Chellappa, R.,
  Zhou, S.K.: Dudonet: Dual domain network for ct metal artifact reduction. In:
  Proceedings of the IEEE Conference on Computer Vision and Pattern
  Recognition. pp. 10512--10521 (2019)

\bibitem{lyu2020encoding}
Lyu, Y., Lin, W.A., Liao, H., Lu, J., Zhou, S.K.: Encoding metal mask
  projection for metal artifact reduction in computed tomography. In:
  International Conference on Medical Image Computing and Computer-Assisted
  Intervention. pp. 147--157. Springer (2020)

\bibitem{meyer2010normalized}
Meyer, E., Raupach, R., Lell, M., Schmidt, B., Kachelrie{\ss}, M.: Normalized
  metal artifact reduction (nmar) in computed tomography. Medical physics
  \textbf{37}(10),  5482--5493 (2010)

\bibitem{ronneberger2015u}
Ronneberger, O., Fischer, P., Brox, T.: U-net: Convolutional networks for
  biomedical image segmentation. In: International Conference on Medical image
  computing and computer-assisted intervention. pp. 234--241. Springer (2015)

\bibitem{wang2018conditional}
Wang, J., Zhao, Y., Noble, J.H., Dawant, B.M.: Conditional generative
  adversarial networks for metal artifact reduction in ct images of the ear.
  In: International Conference on Medical Image Computing and Computer-Assisted
  Intervention. pp. 3--11. Springer (2018)

\bibitem{wang2021dan}
Wang, T., Xia, W., Huang, Y., Sun, H., Liu, Y., Chen, H., Zhou, J., Zhang, Y.:
  Dan-net: Dual-domain adaptive-scaling non-local network for ct metal artifact
  reduction. arXiv preprint arXiv:2102.08003  (2021)

\bibitem{yan2018deep}
Yan, K., Wang, X., Lu, L., Zhang, L., Harrison, A.P., Bagheri, M., Summers,
  R.M.: Deep lesion graphs in the wild: relationship learning and organization
  of significant radiology image findings in a diverse large-scale lesion
  database. In: Proceedings of the IEEE Conference on Computer Vision and
  Pattern Recognition. pp. 9261--9270 (2018)

\bibitem{yu2020deep}
Yu, L., Zhang, Z., Li, X., Xing, L.: Deep sinogram completion with image prior
  for metal artifact reduction in ct images. IEEE Transactions on Medical
  Imaging  \textbf{40}(1),  228--238 (2020)

\bibitem{cnnmar}
{Zhang}, Y., {Yu}, H.: Convolutional neural network based metal artifact
  reduction in x-ray computed tomography. IEEE Transactions on Medical Imaging
  \textbf{37}(6),  1370--1381 (June 2018). \doi{10.1109/TMI.2018.2823083}

\bibitem{zhao2020unsupervised}
Zhao, B., Li, J., Ren, Q., Zhong, Y.: Unsupervised reused convolutional network
  for metal artifact reduction. In: International Conference on Neural
  Information Processing. pp. 589--596. Springer (2020)

\bibitem{zhou2021review}
Zhou, S.K., Greenspan, H., Davatzikos, C., Duncan, J.S., van Ginneken, B.,
  Madabhushi, A., Prince, J.L., Rueckert, D., Summers, R.M.: A review of deep
  learning in medical imaging: Imaging traits, technology trends, case studies
  with progress highlights, and future promises. Proceedings of the IEEE
  (2021)

\bibitem{zhou2019handbook}
Zhou, S.K., Rueckert, D., Fichtinger, G.: Handbook of medical image computing
  and computer assisted intervention. Academic Press (2019)

\end{thebibliography}

\end{document}